%% file: main.tex
\documentclass[times,onecolumn]{aastex62}

\usepackage{cases}
\usepackage{graphicx}
\usepackage{subfigure}
\usepackage{natbib}
\usepackage{color}
\usepackage{tabularx}
\usepackage{bm}
\usepackage{threeparttable}

\newcommand{\thetae}{\theta_{\rm E}}
\newcommand{\pie}{\pi_{\rm E}}

\newcommand{\te}{t_{\rm E}}
\newcommand{\Sp}{{\it Spitzer}}
\newcommand{\event}{OGLE-2015-BLG-1771}

\newcommand{\hjd}{${\rm HJD}^{\prime}$}

\DeclareGraphicsExtensions{.pdf,.png,.jpg}

\shorttitle{}
\shortauthors{Zhang et al.}

\begin{document}

\title{{\large OGLE-2015-BLG-1771Lb: A Microlens Planet Orbiting an Ultracool Dwarf?}}

\correspondingauthor{Weicheng Zang}
\email{zangwc17@mails.tsinghua.edu.cn}

\author{Xiangyu Zhang}
\affiliation{Department of Astronomy and Tsinghua Centre for Astrophysics, Tsinghua University, Beijing 100084, China}

\author[0000-0001-6000-3463]{Weicheng Zang}
\affiliation{Department of Astronomy and Tsinghua Centre for Astrophysics, Tsinghua University, Beijing 100084, China}

\author{Andrzej Udalski}
\affiliation{Warsaw University Observatory, Al. Ujazdowskie 4, 00-478 Warszawa, Poland}

\author{Andrew Gould}
\affiliation{Max-Planck-Institute for Astronomy, K\"onigstuhl 17, 69117 Heidelberg, Germany}
\affiliation{Department of Astronomy, Ohio State University, 140 W. 18th Ave., Columbus, OH 43210, USA}

\author{Yoon-Hyun Ryu}
\affiliation{Korea Astronomy and Space Science Institute, Daejon 34055, Republic of Korea}

\author{Tianshu Wang}
\affiliation{Department of Astronomy and Tsinghua Centre for Astrophysics, Tsinghua University, Beijing 100084, China}
\affiliation{Department of Astrophysical Sciences, Princeton University}

\author{Hongjing Yang}
\affiliation{Department of Astronomy and Tsinghua Centre for Astrophysics, Tsinghua University, Beijing 100084, China}

\author{Shude Mao}
\affiliation{Department of Astronomy and Tsinghua Centre for Astrophysics, Tsinghua University, Beijing 100084, China}
\affiliation{National Astronomical Observatories, Chinese Academy of Sciences, Beijing 100101, China}

\collaboration{(Leading Authors)}


\author{Przemek Mr\'{o}z}
\affiliation{Warsaw University Observatory, Al. Ujazdowskie 4, 00-478 Warszawa, Poland}

\author{Jan~Skowron}
\affiliation{Warsaw University Observatory, Al. Ujazdowskie 4, 00-478 Warszawa, Poland}

\author{Radoslaw~Poleski}
\affiliation{Warsaw University Observatory, Al. Ujazdowskie 4, 00-478 Warszawa, Poland}
\affiliation{Department of Astronomy, Ohio State University, 140 W. 18th Ave., Columbus, OH  43210, USA}

\author{Micha{\l}~K.~Szyma\'{n}ski}
\affiliation{Warsaw University Observatory, Al. Ujazdowskie 4, 00-478 Warszawa, Poland}

\author{Igor Soszy\'{n}ski}
\affiliation{Warsaw University Observatory, Al. Ujazdowskie 4, 00-478 Warszawa, Poland}

\author{Pawe{\l} Pietrukowicz}
\affiliation{Warsaw University Observatory, Al. Ujazdowskie 4, 00-478 Warszawa, Poland}

\author{Szymon Koz{\l}owski}
\affiliation{Warsaw University Observatory, Al. Ujazdowskie 4, 00-478 Warszawa, Poland}

\author{Krzysztof Ulaczyk}
\affiliation{Department of Physics, University of Warwick, Gibbet Hill Road, Coventry, CV4~7AL,~UK}

\collaboration{(THE OGLE COLLABORATION)}


\author{Michael D. Albrow}
\affiliation{University of Canterbury, Department of Physics and Astronomy, Private Bag 4800, Christchurch 8020, New Zealand}

\author{Sun-Ju Chung}
\affiliation{Korea Astronomy and Space Science Institute, Daejon 34055, Republic of Korea}
\affiliation{Korea University of Science and Technology, 217 Gajeong-ro, Yuseong-gu, Daejeon 34113, Republic of Korea}

\author{Cheongho Han}
\affiliation{Department of Physics, Chungbuk National University, Cheongju 28644, Republic of Korea}

\author{Kyu-Ha Hwang}
\affiliation{Korea Astronomy and Space Science Institute, Daejon 34055, Republic of Korea}

\author{Youn Kil Jung}
\affiliation{Center for Astrophysics $|$ Harvard \& Smithsonian, 60 Garden St.,Cambridge, MA 02138, USA}
\affiliation{Korea Astronomy and Space Science Institute, Daejon 34055, Republic of Korea}

\author{In-Gu Shin}
\affiliation{Korea Astronomy and Space Science Institute, Daejon 34055, Republic of Korea}

\author{Yossi Shvartzvald}
\affiliation{Department of Particle Physics and Astrophysics, Weizmann Institute of Science, Rehovot 76100, Israel}

\author{Jennifer~C.~Yee}
\affiliation{Center for Astrophysics | Harvard \& Smithsonian, 60 Garden St.,Cambridge, MA 02138, USA}

\author{Wei Zhu}
\affiliation{Canadian Institute for Theoretical Astrophysics, University of Toronto, 60 St George Street, Toronto, ON M5S 3H8, Canada}

\author{Sang-Mok Cha}
\affiliation{Korea Astronomy and Space Science Institute, Daejon 34055, Republic of Korea}
\affiliation{School of Space Research, Kyung Hee University, Yongin, Kyeonggi 17104, Republic of Korea} 

\author{Dong-Jin Kim}
\affiliation{Korea Astronomy and Space Science Institute, Daejon 34055, Republic of Korea}

\author{Hyoun-Woo Kim}
\affiliation{Korea Astronomy and Space Science Institute, Daejon 34055, Republic of Korea}
\affiliation{Department of Astronomy and Space Science, Chungbuk National University, Cheongju 28644, Republic of Korea}

\author{Seung-Lee Kim}
\affiliation{Korea Astronomy and Space Science Institute, Daejon 34055, Republic of Korea}
\affiliation{Korea University of Science and Technology, 217 Gajeong-ro, Yuseong-gu, Daejeon 34113, Republic of Korea}

\author{Chung-Uk Lee}
\affiliation{Korea Astronomy and Space Science Institute, Daejon 34055, Republic of Korea}
\affiliation{University of Science and Technology, Korea, (UST), 217 Gajeong-ro Yuseong-gu, Daejeon 34113, Republic of Korea}

\author{Dong-Joo Lee}
\affiliation{Korea Astronomy and Space Science Institute, Daejon 34055, Republic of Korea}

\author{Yongseok Lee}
\affiliation{Korea Astronomy and Space Science Institute, Daejon 34055, Republic of Korea}
\affiliation{School of Space Research, Kyung Hee University, Yongin, Kyeonggi 17104, Republic of Korea}

\author{Byeong-Gon Park}
\affiliation{Korea Astronomy and Space Science Institute, Daejon 34055, Republic of Korea}
\affiliation{Korea University of Science and Technology, 217 Gajeong-ro, Yuseong-gu, Daejeon 34113, Republic of Korea}

\author{Richard W. Pogge}
\affiliation{Department of Astronomy, Ohio State University, 140 W. 18th Ave., Columbus, OH 43210, USA}
\collaboration{(The KMTNet Collaboration)}

\input{abstract}

\section{Introduction}\label{intro}
\input{intro}

\section{Observations}\label{obser}
\input{obser}

\section{Light curve analysis}\label{model}

\input{model}

\section{Physical Parameters}\label{lens}
\input{lens}

\section{Discussion}\label{dis}

\input{dis}

\bibliography{Zang.bib}

\input{table.tex}
\input{figure.tex}

\end{document}

%% file: abstract.tex
\begin{abstract}
We report the discovery and the analysis of the short ($\te < 5$~days) planetary microlensing event, \event. The event was discovered by the Optical Gravitational Lensing Experiment (OGLE), and the planetary anomaly (at $I \sim 19$) was captured by The Korea Microlensing Telescope Network (KMTNet). The event has three surviving planetary models that explain the observed light curves, with planet-host mass ratio $q \sim 5.4 \times 10^{-3}, 4.5 \times 10^{-3} $ and $4.5 \times 10^{-2}$, respectively. The first model is the best-fit model, while the second model is disfavored by $\Delta\chi^2 \sim 3$. The last model is strongly disfavored by $\Delta\chi^2 \sim 15$ but not ruled out. A Bayesian analysis using a Galactic model indicates that the first two models are probably composed of a Saturn-mass planet orbiting a late M dwarf, while the third one could consist of a super-Jovian planet and a mid-mass brown dwarf. The source-lens relative proper motion is $\mu_{\rm rel} \sim 9~{\rm mas\,yr^{-1}}$, so the source and lens could be resolved by current adaptive-optics (AO) instruments in 2021 if the lens is luminous.
\end{abstract}

%% file: intro.tex
Early observations using ALMA \citep{Testi2016} and {\it Herschel} \citep{Daemgen2016} suggest that disks around ultracool dwarfs are frequent. Searching for and studying planets around ultracool dwarfs are important for the conditions for planet formation theories \citep[e.g.,][]{Ida2005,Boss2006} at the low-mass end. However, the detection of planets around ultracool dwarfs is challenging due to the intrinsic faintness of the host stars. At the time of writing, more than 4000 confirmed exoplanets have been detected\footnote{\url{http://exoplanetarchive.ipac.caltech.edu} as of 2019 October 31.}, but only 21 of them are orbiting a $M_{\rm host} < 0.1M_{\odot}$ star. 

Among the 21 such known planets, four of them were found by direct imaging method: 2MASS 1207-3932 \citep{2004A&A...425L..29C}, 2MASS 0441-2301 \citep{2010ApJ...714L..84T}, VHS 1256-1257 \citep{2015ApJ...804...96G}, CFBDSIR 1458+1013 \citep{2011ApJ...740..108L}. All of these planets are super-Jovian planets ($> 4M_J$) and have a planet-host mass ratio $q > 0.15$, which indicates that these systems may form similarly to binary systems. In addition, seven temperate terrestrial planets were discovered around the nearby ultracool dwarf stars TRAPPIST-1 \citep{TRAPPIST-1} via the transit method, and two similar planets around Teegarden's Star were detected by the radial velocity method \citep{Teegarden}, which suggests that terrestrial planets should be frequent around ultracool dwarfs. 

Microlensing opens a powerful window for probing planets around ultracool dwarfs because it does not rely on the light from the host stars but rather uses the light from a background source \citep{Shude1991,Andy1992}. Microlensing has detected three planets orbiting a $M_{\rm host} < 0.1M_{\odot}$ star with unambiguous mass measurements. \cite{OB161195_MOA} and \cite{OB161195} detected a $q \sim 6\times10^{-5}$ planet in the micolensing event OGLE-2016-BLG-1195, and a joint analysis of ground-based and \Sp\ data \citep{OB161195} revealed that this planetary system is composed of an Earth-mass ($\sim1.4M_{\earth}$) planet around a $\sim0.078M_{\odot}$ ultracool dwarf. \cite{OB120358} discovered a $\sim2M_{J}$ planet orbiting a $\sim0.02M_{\odot}$ very low mass brown dwarf (BD) in the event OGLE-2012-BLG-0358, and \cite{MB07192} detected a $\sim3M_{\earth}$ super-Earth planet around a $\sim0.08M_{\odot}$ ultracool dwarf \citep{MB07192_AO} in the event MOA-2007-BLG-192. For the planets using Bayesian analysis to estimate the host mass, \cite{KB161820} reported a super-Jovian planet orbiting a $M_{\rm host} < 0.1M_{\odot}$ star with a $\sim90\%$ probability. \cite{OB171522} reported a Jovian-mass planet around a BD, but the host star also has a $\sim30\%$ probability to be a M-dwarf or K-dwarf. In addition, there are three events with degenerate solutions. Bayesian analysis shows that one of the solution of MOA-2015-BLG-337 \citep{MB15337} and KMT-2016-BLG-1107 \citep{KB161107} probably consist of a giant planet orbiting a BD. \cite{MB13605} found three degenerate planetary models in the event MOA-2013-BLG-605, two of which suggest a super-Earth orbiting a BD. For the five events using Bayesian analysis to estimate the host mass and/or that have degenerate solutions, we can verify that the host is an ultracool dwarf by adaptive-optics (AO) instruments in the future.

Here we report the analysis of the microlens planetary event \event. The observed data are consistent with three planetary models, and a Bayesian analysis suggests the host star is likely an ultracool dwarf ($M_{\rm host} < 0.2M_{\odot}$). The paper is structured as follows. In Section \ref{obser}, we introduce data acquisition and processing of this event. We then describe the light curve analysis in Section \ref{model} and estimate the physical parameters of the planetary system in Section \ref{lens}. Finally, we discuss the implications of our work in Section \ref{dis}.

%% file: obser.tex
\event\ was discovered by the Optical Gravitational Lensing Experiment (OGLE, \citealt{OGLEIV}) using its 1.3 m Warsaw Telescope at the Las Campanas Observatory in Chile and alerted by the OGLE Early Warning System \citep{Udalski1994,Udalski2003} at UT 00:46 on 2 August 2015. The event was located at equatorial coordinates $(\alpha, \delta)_{\rm J2000}$ = (17:55:11.76, $-28$:51:45.9), corresponding to Galactic coordinates $(\ell,b)=(1.14, -1.76)$. It therefore lies in OGLE field BLG505, monitored by OGLE with a cadence of \textbf{$\Gamma = 3\,{\rm hr}^{-1}$}. The event was also observed by the Korea Microlensing Telescope Network (KMTNet, \citealt{KMT2016}). KMTNet consists of three 1.6~m telescopes, equipped with 4 ${\rm deg}^2$ FOV cameras at the Cerro Tololo International Observatory (CTIO) in Chile (KMTC), the South African Astronomical Observatory (SAAO) in South Africa (KMTS), and the Siding Spring Observatory (SSO) in Australia (KMTA). The event was located in the KMTNet BLG02 field, which was observed in 2015 with a cadence of $\Gamma = 6\,{\rm hr}^{-1}$. The majority of observations by OGLE and KMTNet were taken in the $I$-band, with some $V$-band images taken for the color measurement of microlens sources. However, the $V$-band data have signal-to-noise ratio (SNR) too low to determine the source color. The photometry of OGLE and KMTNet was extracted using custom implementations of the difference image analysis technique \citep{Alard1998}: \citealt{Wozniak2000} (OGLE) and \citealt{pysis} (KMTNet).

%% file: model.tex
Figure \ref{lc1} shows the observed data together with the best-fit models. The light curve shows a ``U'' shape bump at \hjd~$\sim 7235.1 ({\rm HJD}^{\prime} = {\rm HJD} - 2450000$), which is generally produced by a caustic crossing in a binary-lensing (2L1S) event, so we fit the data with the 2L1S model in Section \ref{2L1S}. We also check the binary-source (1L2S) model in Section \ref{1L2S}.

\subsection{Binary-Lens Model}\label{2L1S}
Standard binary lens models require seven parameters to calculate the magnification, $A(t)$. The first three are point-lens parameters ($t_0$, $u_0$, $\te$) \citep{Paczynski1986}: the time at which the source passes closest to the center of lens mass, the impact parameter normalized by the angular Einstein radius $\thetae$, and the Einstein radius crossing time, respectively. The next three ($q$, $s$, $\alpha$) define the binary companion: the mass ratio, the projected separation between the binary components scaled to $\thetae$, and the angle between the source trajectory and the binary axis in the lens plane, respectively. The last one $\rho$ is the angular source radius $\theta_*$ scaled to $\thetae$ ($\rho = \theta_*/\thetae$). We use the advanced contour integration code \citep{Bozza2010}, \texttt{VBBinaryLensing}\footnote{\url{http://www.fisica.unisa.it/GravitationAstrophysics/VBBinaryLensing.htm}}, to compute the binary-lens magnification $A(t)$. In addition, for each data set $i$, we introduce two flux parameters ($f_{{\rm S},i}$, $f_{{\rm B},i}$) to represent the flux of the source star and any additional blend flux. The observed flux, $f_{i}(t)$, calculated from the model, is

\begin{equation}
    f_{i}(t) = f_{{\rm S},i} A(t) + f_{{\rm B},i}.
\end{equation}

We follow the method of \cite{KB161836} to search for the best-fit models. In brief, we initially conduct a sparse grid search over ($\log s, \log q, \alpha, \rho$) to roughly locate the solutions in the ($\log s, \log q$) plane, and then undertake a denser grid search over ($\log s, \log q$) on those promising locations to find the $\chi^2$ minima. Finally, setting the initial parameters as those minima, we investigate the best-fit model with all free parameters by Markov chain Monte Carlo (MCMC) $\chi^2$ minimization using the \texttt{emcee} ensemble sampler \citep{emcee}.

As shown in Figure \ref{grid}, we find five distinct minima (labeled as ``A'', ``B'', ``C'', ``D'' and ``E'' in the lower panel Figure \ref{grid}). The MCMC results show that the Model ``A'' provides the best fit to the observed data, while the Models ``B'', ``C'', ``D'' and ``E'' are disfavored by $\Delta\chi^2 \sim 3, 15, 54~{\rm and}~134$, respectively (see Table \ref{parm} for the parameters). Figure \ref{cau} shows the lens-system configurations of the individual 
degenerate models. In Figures \ref{lc2} and \ref{chi2}, we find that most of the $\chi^2$ difference of Models ``D'' and ``E'' are from the anomalous region. Together with the relatively large $\Delta\chi^2$, we only investigate Models  ``A'', ``B'' and ``C'' in the following analysis. In addition, all the surviving models (A, B, and C) have very low mass ratios, indicating that the companion is a planetary-mass object.

In some cases, the microlens parallax $\pie$ can be measured by considering the orbital motion of Earth around the Sun in the light curve analysis \citep{Gould1992, Alcock1995}. However, this method is generally feasible only for events with long timescale $\te\gtrsim$ year/$2\pi$ \citep[e.g.,][]{OB171434} that introduce significant deviation from rectilinear motion in the lens-source relative motion. For \event, the timescale $\te < 5$~days, so the parallax effect should be negligible. As a result, the addition of parallax to the models only provides $\Delta\chi^2 < 2$, and the upper limit of the microlens parallax as the $1\sigma$ level is $\pie \la 100$ for all the three models, which gives no useful constraint.

\subsection{Binary-Source Model}\label{1L2S}
A binary-source event is the superposition of two point-lens events. \cite{Gaudi1998} first pointed out that a 1L2S event can mimic a 2L1S event if the binary source (labeled as ``S1'' and ``S2'') has a large flux ratio $q_F = f_{\rm S1}/f_{\rm S2}$ and the second source ``S2'' pass much closer to the lens. We therefore search for 1L2S solutions using MCMC, which shows that the best-fit 1L2S model is disfavored by $\Delta\chi^2 \sim 86$ compared to the best-fit 2L1S model (see Table \ref{parm} for the parameters). In Figure \ref{chi2}, we find that most of the $\chi^2$ difference comes from the anomalous region, in which the 1L2S model cannot fit the ``U'' shape of the anomalous region. Thus, we exclude the 1L2S solution.

%% file: lens.tex
Uniquely determining the total lens mass $M_{\rm L}$ and distance $D_{\rm L}$ requires two observables: the angular Einstein radius $\thetae$ and the microlens parallax $\pie$ \citep{Gould1992, Gould2000}
\begin{equation}
    M_{\rm L} = \frac{\thetae}{{\kappa}\pie}, \qquad D_{\rm L} = \frac{\mathrm{AU}}{\pie\thetae + \pi_{\rm S}},
\end{equation}
where $\kappa \equiv 4G/(c^2\mathrm{AU}) = 8.144$ mas$/M_{\odot}$, $\pi_{\rm S} = \mathrm{AU}/D_{\rm S}$ is the source parallax, and $D_{\rm S}$ is the source distance. We estimate the angular Einstein radius by $\thetae = \theta_*/\rho$ in Section \ref{CMD}. However, the observed data give no useful constraint on the microlens parallax (see Section \ref{2L1S}). Thus, we conduct a Bayesian analysis in Section \ref{Bayesian} to estimate the physical parameters of the planetary system. 

\subsection{Color Magnitude Diagram}\label{CMD}
We estimate the angular source radius $\theta_*$ based on the de-reddened brightness and color of the source \citep{Yoo2004}. We construct the color magnitude diagram (CMD) using OGLE stars within a $2' \times 2'$ square centered on the position of the event (see Figure \ref{cmd}). We measure the centroid of the red giant clump as $(V - I, I)_{\rm cl} = (2.65 \pm 0.01, 16.68 \pm 0.01)$, and compare it to the intrinsic centroid of the red giant clump $(V - I, I)_{\rm cl,0} = (1.06, 14.39)$ \citep{Bensby2013,Nataf2016}, which yields an offset $\Delta(V - I, I)_{\rm cl} = (1.59 \pm 0.02, 2.29\pm 0.03)$. 

From the light curve modeling, the source apparent brightness is $I_{\rm S, A} = 21.77 \pm 0.08$, $I_{\rm S, B} = 21.86 \pm 0.06$ and $I_{\rm S, C} = 20.91 \pm 0.05$ for Models ``A'', ``B'' and ``C'', respectively. However, in this case we have no color measurements of the source due to too low signal-to-noise in $V$-band. Nevertheless, it is still possible to estimate the source color following the method of \cite{MB07192} and \cite{MOAbin29}. We first calibrate the CMD of \cite{HSTCMD} HST observations to the OGLE CMD using its red-clump centroid of $(V - I, I)_{\rm cl, HST}=(1.62, 15.15)$ \citep{MB07192}. We then estimate the source color by taking the average color of the calibrated Holtzman field stars whose brightness are within the $3\sigma$ of the microlens source star. Using the derived offset of the red giant clump, the de-reddened brightness $I_{\rm S,0}$ and color $(V - I)_{\rm S,0}$ of the source can be measured. Finally, we apply the color/surface-brightness relation of \cite{Adams2018} to estimate the angular source radius $\theta_*$. We summarize the values of the source and the derived angular Einstein radius $\thetae$ and the lens-source relative proper motion $\mu_{\rm rel}$ in Table \ref{source}.

\subsection{Bayesian Analysis}\label{Bayesian}
Our Bayesian analysis is based on the Galactic model of \cite{OB171522} derived from the models of \cite{Han1995} and \cite{Han2003}. Because the timescale of the event is $< 5$ days, we expect that objects in the planetary mass regime are also plausible lenses \citep[e.g.,][]{MB15337}. We therefore adopt a broken power-law mass function as follows,

\begin{numcases}{dN/dM=}
a_0M^{-\alpha_{\rm pl}} \qquad (0.001 \leq M/M_{\odot} \leq 0.013) \\
a_1M^{-0.3} \qquad (0.013 \leq M/M_{\odot} \leq 0.08) \\
a_2M^{-1.3} \qquad (0.08 \leq M/M_{\odot} \leq 0.5) \\
a_3M^{-2.3} \qquad (0.5 \leq M/M_{\odot} \leq 1.3)
\end{numcases}
where the last three terms are the Kroupa mass function \citep{Kroupa2001} used in \cite{Zhu2017spitzer}, ($a_0, a_1, a_2, a_3$) are normalizing coefficients, and $\alpha_{\rm pl}$ is the slope of the planetary mass regime. We create a sample of $10^9$ simulated events for $\alpha_{\rm pl} = -4.0~{\rm and}~0.6$, respectively. The planetary slope $\alpha_{\rm pl} = -4.0$ is similar to that of \cite{Mroz2017a} for unbound or wide-orbit Jupiter-mass planets. $\alpha_{\rm pl} = 0.6$ has $1 : 0.26$ for the relative fractions of number between main sequence stars and planetary mass objects, which is just slightly higher than the result of \cite{Mroz2017a} who found that the upper limit on the frequency of Jupiter-mass free-floating or wide-orbit planets is 0.25 per main sequence star at $95\%$ confidence. For each simulated event $i$ of model $k$, the weight is given by
\begin{equation}
    W_{{\rm Gal},i,k} = \Gamma_{i,k} \mathcal{L}_{i,k}(\te) \mathcal{L}_{i,k}(\thetae),
\end{equation}
where $\Gamma_{i,k}\varpropto\theta_{{\rm E},i,k}\times\mu_{{\rm rel},i,k}$ is the microlensing event rate, $\mathcal{L}_{i,k}(\te)$ and $\mathcal{L}_{i,k}(\thetae)$ are the likelihood of its derived parameters $(\te, \thetae)_{i,k}$ given the error distributions of these quantities for that model
\begin{equation}
    \mathcal{L}_{i,k}(X) = \frac{{\rm exp}[-(X_{i} - X_{k})^2/2\sigma^2_{X_{k}}]}{\sqrt{2\pi}\sigma_{X_{k}}}, \qquad X = \te~{\rm or}~X = \thetae.
\end{equation}

The resulting posterior distributions of the lens host-mass $M_{\rm host}$, the lens distance $D_{\rm L}$, the planet mass $M_{\rm planet}$, the projected planet-host separation $r_{\bot}$, the angular Einstein radius $\thetae$ and the lens-source relative proper motion $\mu_{\rm rel}$ for Models ``A'', ``B'' and ``C'' are shown Figure \ref{fig:baye} and Table \ref{table:baye}. For Models ``A'' and ``B'', the effects of different $\alpha_{\rm pl}$ are negligible, and the planetary system is probably composed of a Saturn-mass planet orbiting a late M dwarf. For Model ``C'', the distributions of planetary host mass ($M_{\rm host} < 13M_{J}$) are different for the two $\alpha_{\rm pl}$, with $3.2\%$ probability distribution for  $\alpha_{\rm pl} = -4.0$ and $12.0\%$ for $\alpha_{\rm pl} = 0.6$. Because both distributions indicate a mid-mass BD host star, we adopt the distributions of $\alpha_{\rm pl} = -4.0$ for the final lens properties. The projected planet-host separation is $\sim 0.5$--$1.0$ AU for the three models, which indicates that the planet is well beyond the snow line (assuming a snow line radius $r_{\rm SL} = 2.7(M/M_{\odot})$~{\rm AU}, \citealt{snowline}).


%% file: dis.tex
We have reported the discovery and analysis of the microlens planet OGLE-2015-BLG-1771Lb. Our analysis suggests that the planetary system probably consists of a gas giant planet and an ultracool dwarf. This conclusion is based on a Bayesian analysis that shows the lens has a $\sim 65\%$ probability of being $<0.1 M_\odot$ and a $\sim 85\%$ probability of being $<0.2 M_\odot$ (for $\alpha_{\rm pl} = -4.0$). Of course, this still leaves a significant possibility that the lens could be a more massive star. For example, similar to this event, the Bayesian posterior for the primary of OGLE-2014-BLG-0962 \citep{OB140962} peaks at a mass of $\sim 0.07 M_{\odot}$ with an $84\%$ probability that the mass is $<0.2 M_{\odot}$. However, including the parallax measurement for that event yields a measured mass of $0.2 M_{\odot}$. In the present case, we can verify within a few years that the host is an ultracool dwarf by excluding stellar mass hosts for \event\ with high-resolution imaging. The measured source-lens relative proper motion for the three models is quite large (see Table \ref{source}) and the source is quite faint ($I > 20.7$). This is similar to the case of OGLE-2005-BLG-169 for which HST (\citep{Dave2015}) was able to resolve the source and the lens when they were separated by $\sim 48$~mas and Keck adaptive optics (\citep{Batista2015}) resolved them at a separation $\sim 60$~mas. Thus, even for model A (which has the lowest proper motion, $\mu_{\rm rel} \sim 8.5~{\rm mas\,yr^{-1}}$), the source and lens will be separated by $\sim 60$~mas as soon as 2021. Because the source is faint ($I>20.7$), we can expect stringent constraints on the lens light if it is not luminous.

For many years (beginning with the second microlens planet, OGLE-2005-BLG-071Lb, \citealt{OB050071}), most microlensing planets were discovered based on the strategy advocated by \cite{Andy1992} using a combination of wide-area surveys for finding microlensing events and intensive follow-up observations for capturing the planetary perturbation. The second generation microlensing surveys, conducted by The Microlensing Observations in Astrophysics (MOA, \citealt{MOA2016}), OGLE, Wise Observatory \citep{Wise} and KMTNet, aim to detect planets by wide-area, high-cadence observations, without the need for follow-up observations. For the planet OGLE-2015-BLG-1771Lb, the event timescale ($< 5$~days) and the planetary signal ($\sim 5$~hours) are short, and the anomaly is faint ($I_{\rm anom} \sim 19$), so the planet can only be detected by second generation microlensing surveys. For those nine microlens planets which have a $>50\%$ probability to orbit a $M_{\rm host} < 0.1M_{\odot}$ host star, only OGLE-2012-BLG-0358Lb was detected using the strategy of \cite{Andy1992}. Moreover, the rate of discovery such planets is much higher beginning with 2015 (i.e., the observations of KMTNet), during which 6/9 planets were detected. In addition, the typical time scale $\te$ for the microlensing events with a $M < 0.1M_{\odot}$ lens is $\lesssim10$~days. For the three planets detected before 2015, all of them have $\te > 20$ days, while 5/6 planets beginning with 2015 have $\te < 10$ days, which suggests that the current second generation microlensing surveys are more sensitive to the planets around ultracool dwarfs. Future statistical analyses of the microlens sample of planets around untracool dwarfs will potentially reveal the properties of such planets and thus provide stringent constraints on the planet formation theories.

\acknowledgments
We thank Chris W. Ormel and Xuening Bai for fruitful discussions. X.Z., W.Z., W.T., H.Y. and S.M. acknowledge support by the National Science Foundation of China (Grant No. 11821303 and 11761131004). The OGLE has received funding from the National Science Centre, Poland, grant MAESTRO 2014/14/A/ST9/00121 to AU. Work by AG was supported by AST-1516842 and by JPL grant 1500811. This research has made use of the KMTNet system operated by the Korea Astronomy and Space Science Institute (KASI) and the data were obtained at three host sites of CTIO in Chile, SAAO in South Africa, and SSO in Australia. AG received support from the European Research Council under the European Unions Seventh Framework Programme (FP 7) ERC Grant Agreement n. [321035]. Work by CH was supported by the grants (2017R1A4A1015178 and 2019R1A2C2085965) of National Research Foundation of Korea. Wei Zhu was supported by the Beatrice and Vincent Tremaine Fellowship at CITA. This research has made use of the NASA Exoplanet Archive, which is operated by the California Institute of Technology, under contract with the National Aeronautics and Space Administration under the Exoplanet Exploration Program.

%% file: table.tex
\begin{table}[htb]
    \centering
    \caption{Best-fit models and their $68\%$ uncertainty range from MCMC}
    \begin{tabular}{c c c c c c c}
    \hline
    \hline
    Models  & A & B & C & D & E & Binary source \\
    \hline
    $t_{0,1}$ (${\rm HJD}^{\prime}$) & $7235.60 \pm 0.01$ & $7235.62 \pm 0.01$ & $7235.51 \pm 0.01$ & $7235.61 \pm 0.01$ & $7234.74 \pm 0.23$ & $7235.77 \pm 0.02$ \\
    $t_{0,2}$ (${\rm HJD}^{\prime}$) & ... & ... & ... & ... & ... & $7235.06 \pm 0.03$ \\
    $u_{0,1}$  & $0.121 \pm 0.008$ & $0.114 \pm 0.006$ & $0.242 \pm 0.009$ & $0.273 \pm 0.016$ & $0.024 \pm 0.016$ & $0.112 \pm 0.024$ \\
    $u_{0,2}$  & ... &... & ... & ... & ... & $0.001 \pm 0.025$ \\
    $\te$ (days)  & $4.28 \pm 0.24$ & $4.53 \pm 0.18$ & $2.49 \pm 0.10$ & $2.62 \pm 0.12$ & $8.64 \pm 0.98$ & $5.39 \pm 0.85$ \\
    $s$ & $1.202 \pm 0.010$ & $0.998 \pm 0.008$ & $1.119 \pm 0.006$ & $0.850 \pm 0.008$ & $2.216 \pm 0.090$ & ... \\
    $q (10^{-3})$  & $5.38 \pm 0.64$ & $4.47 \pm 0.51$ & $45.5 \pm 4.5$ & $3.39 \pm 0.35$ & $70.9 \pm 9.8$ & ...\\
    $\alpha$ (deg) & $223.7 \pm 1.2$ & $222.4 \pm 0.4$ & $191.9 \pm 0.4$ & $38.6 \pm 0.4$ & $146.8 \pm 0.8$ & ...\\
    $\rho_1(10^{-3})$ & $4.41 \pm 0.46$ & $3.64 \pm 0.34$  & $8.27 \pm 0.80$ & $9.15 \pm 0.78$ & $4.41 \pm 0.50$ & $131 \pm 27$ \\
    $\rho_2(10^{-3})$ & ... & ... & ... & ... & ... & $10 \pm 2$ \\
    $q_{F}$ & ... & ... & ... & ... & ... & $0.080 \pm 0.012$ \\
    $I_{\rm S}$ & $21.77 \pm 0.08$ & $21.86 \pm 0.06$ & $20.91 \pm 0.05$ & $20.82 \pm 0.08$ & $22.86 \pm 0.13$ & $22.25 \pm 0.24$ \\
    $I_{\rm B}$ & $21.03 \pm 0.04$ & $20.99 \pm 0.03$  & $22.04 \pm 0.14$ & $22.36 \pm 0.30$ & $20.74 \pm 0.02$ & $20.85 \pm 0.06$ \\
    $\chi^2/dof$ &  $3489.8/3481$ & $3492.7/3481$ & $3505.1/3481$ & $3543.5/3481$ & $3624.1/3481$ & $3575.4/3480$ \\
    \hline
    \hline
    \end{tabular}
    \label{parm}
\end{table}

\begin{table}[htb]
    \renewcommand\arraystretch{1.5}
    \centering
    \caption{The de-reddened source color and magnitude, the values of $\theta_*$, $\thetae$ and $\mu_{\rm rel}$}
    \begin{tabular}{c c c c c}
    \hline
    \hline
    Models & Unit & A & B & C \\
    \hline
    $I_{\rm S,0}$  & mag & $19.48 \pm 0.09$ & $19.57 \pm 0.07$ & $18.62 \pm 0.06$  \\
    $(V - I)_{\rm S,0}$  & mag & $0.92 \pm 0.14$ & $0.95 \pm 0.16$ & $0.78 \pm 0.09$  \\
    $\theta_*$ & $\mu{\rm as}$  & $0.49 \pm 0.08$  & $0.48 \pm 0.08$  & $0.65 \pm 0.07$ \\
    \hline
    $\thetae$ & mas & $0.111 \pm 0.022$ &  $0.132 \pm 0.025$ & $0.079 \pm 0.011$   \\ 
    $\mu_{\rm rel}$ & ${\rm mas\,yr^{-1}}$ & $9.5 \pm 2.0$ & $10.6 \pm 2.1$ & $11.6 \pm 1.7$ \\
    \hline
    \hline
    \end{tabular}
    \label{source}
\end{table}

\begin{table}[htb]
    \renewcommand\arraystretch{1.5}
    \centering
    \caption{Physical parameters for \event}
    \begin{tabular}{c c | c c c | c c c}
    \hline
    \hline
    & & \multicolumn{3}{c|}{$\alpha_{\rm pl} = -4.0$}  & \multicolumn{3}{c}{$\alpha_{\rm pl} = 0.6$}  \\
    \hline
    Lens Parameters & Unit  & Model A & Model B & Model C & Model A &  Model B  & Model C\\
    \hline
    $M_{\rm host}$ & $M_{\odot}$ & $0.077^{+0.119}_{-0.044}$ & $0.086^{+0.133}_{-0.047}$ & $0.055^{+0.091}_{-0.033}$ & $0.076^{+0.119}_{-0.044}$ & $0.085^{+0.132}_{-0.047}$ & $0.049^{+0.091}_{-0.033}$ \\
    $M_{\rm planet}$ & $M_{J}$ & $0.433^{+0.674}_{-0.251}$ & $0.401^{+0.624}_{-0.226}$ & $2.634^{+4.361}_{-1.615}$ & $0.427^{+0.672}_{-0.255}$ & $0.397^{+0.620}_{-0.227}$ & $2.331^{+4.368}_{-1.576}$\\
    $D_{\rm L}$ & kpc & $7.07^{+1.00}_{-1.09}$ & $6.86^{+1.04}_{-1.14}$ & $6.96^{+0.96}_{-1.00}$ & $7.04^{+1.02}_{-1.14}$ & $6.83^{+1.05}_{-1.17}$ & $6.85^{+1.02}_{-1.15}$ \\
    $r_{\bot}$ & AU &$0.85^{+0.16}_{-0.16}$ & $0.78^{+0.15}_{-0.15}$ & $0.56^{+0.09}_{-0.08}$ & $0.85^{+0.16}_{-0.17}$ & $0.77^{+0.15}_{-0.15}$ &  $0.55^{+0.09}_{-0.10}$ \\
    $\thetae$ &  mas  & $0.100^{+0.019}_{-0.018}$ & $0.114^{+0.021}_{-0.020}$ & $0.072^{+0.010}_{-0.010}$    &  $0.100^{+0.019}_{-0.018}$   &  $0.115^{+0.021}_{-0.021}$    &  $0.071^{+0.010}_{-0.010}$ \\
    $\mu_{\rm rel}$ & ${\rm mas\,yr^{-1}}$ & $8.5^{+1.6}_{-1.5}$ & $9.2^{+1.7}_{-1.6}$  & $10.4^{+1.4}_{-1.4}$    &  $8.5^{+1.6}_{-1.5}$  &   $9.2^{+1.7}_{-1.6}$    & $10.4^{+1.4}_{-1.4}$  \\
    \hline
    \hline
    \end{tabular}
    \label{table:baye}
\end{table}

%% file: figure.tex
\begin{figure}[htb] 
    \includegraphics[width=\columnwidth]{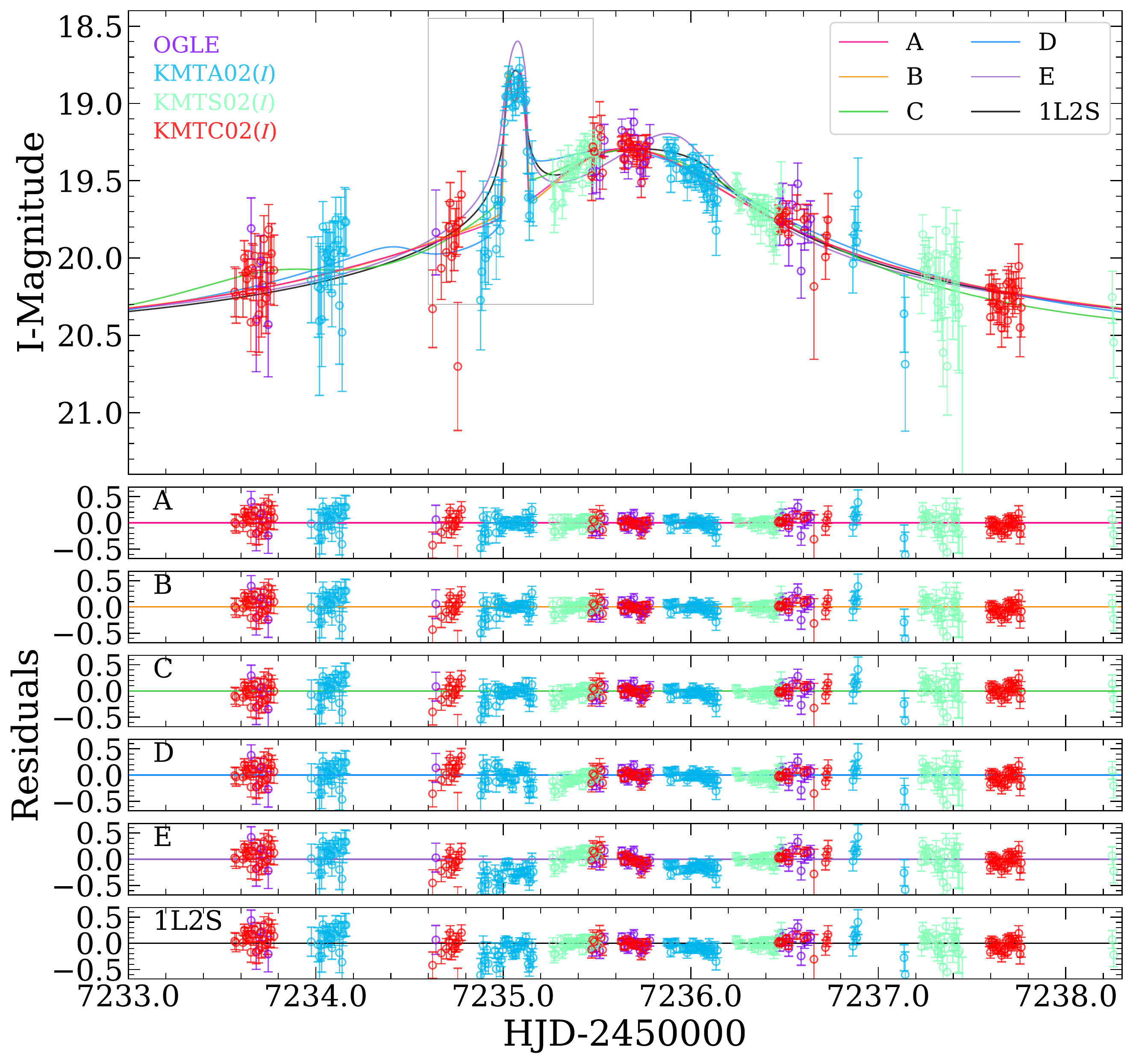}
    \caption{The data of \event\ together with the best-fit 2L1S and 1L2S models. Data points for different data set and light curves for different models are shown with different colors.}
    \label{lc1}
\end{figure}

\begin{figure*}[htb]
    \centering
    \subfigure{
    \begin{minipage}{15cm}
    \centering
    \includegraphics[width=\columnwidth]{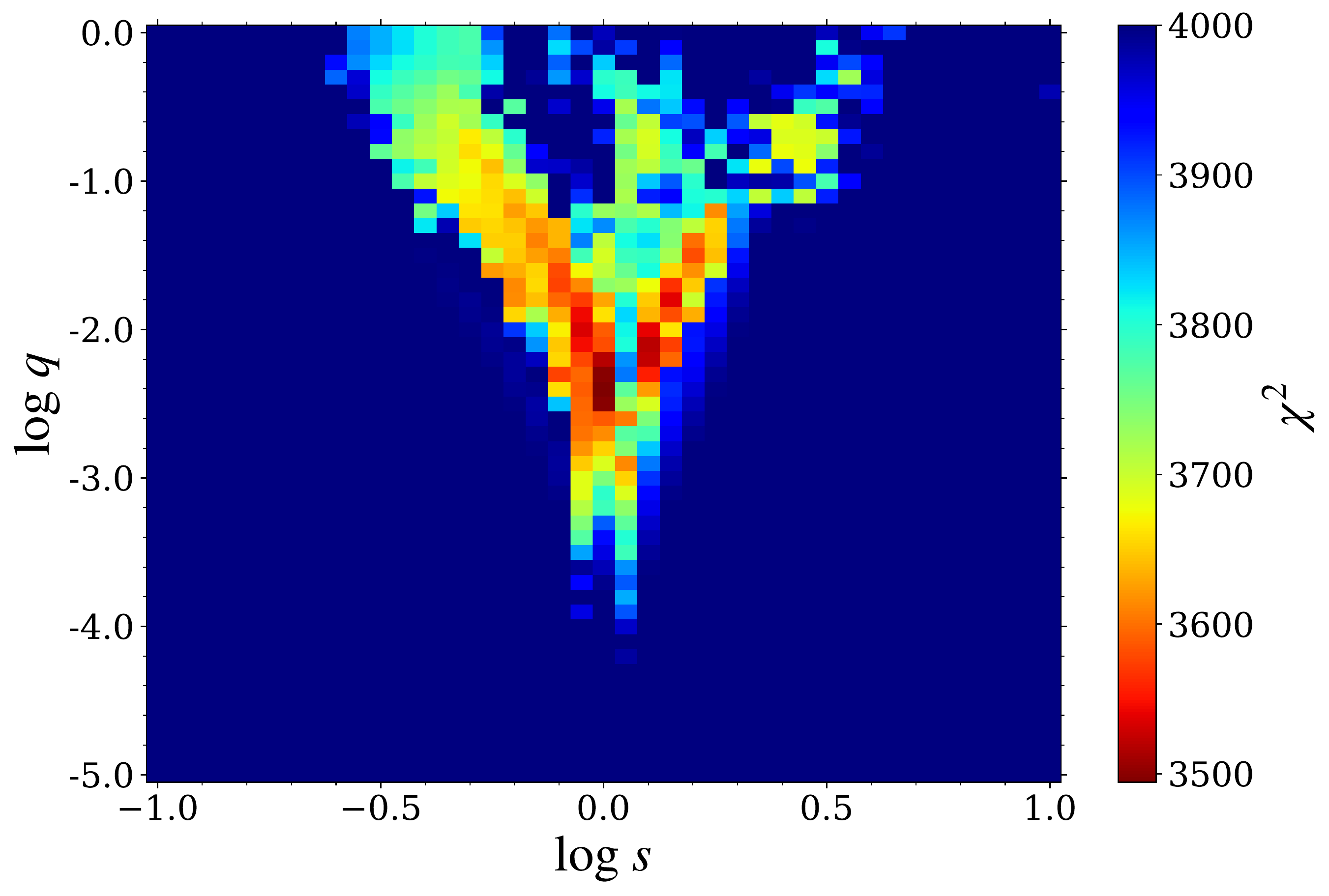}
    \end{minipage}
    }
    \subfigure{
    \begin{minipage}{15cm}
    \centering
    \includegraphics[width=\columnwidth]{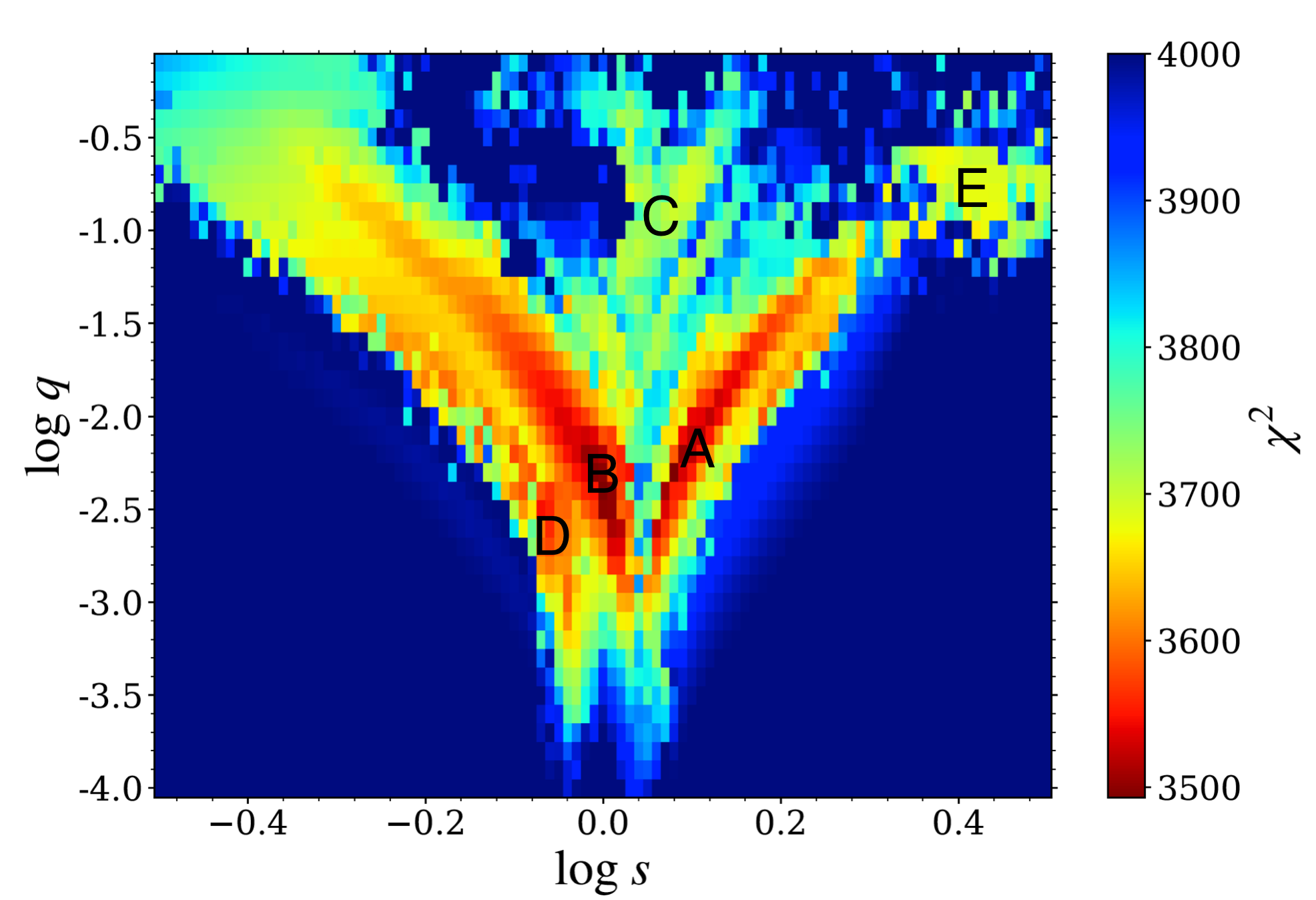}
    \end{minipage}
    }
    \caption{$\chi^2$ distributions of the grid search projected onto the ($\log s, \log q$) plane. The upper panel shows the space that is equally divided on a ($41 \times 51$) grid with ranges of $-1.0\leq\log s \leq1.0$ and $-5.0\leq \log q \leq0$, respectively. The lower panel shows the space that is equally divided on a ($101 \times 41$) grid with ranges of $-0.5\leq\log s \leq0.5$ and $-4.0\leq \log q \leq0.0$, respectively. The labels ``A'', ``B'', ``C'', ``D'' and ``E'' in the lower panel represent five distinct minima.}
    \label{grid}
\end{figure*}

\begin{figure}[htb] 
    \includegraphics[width=\columnwidth]{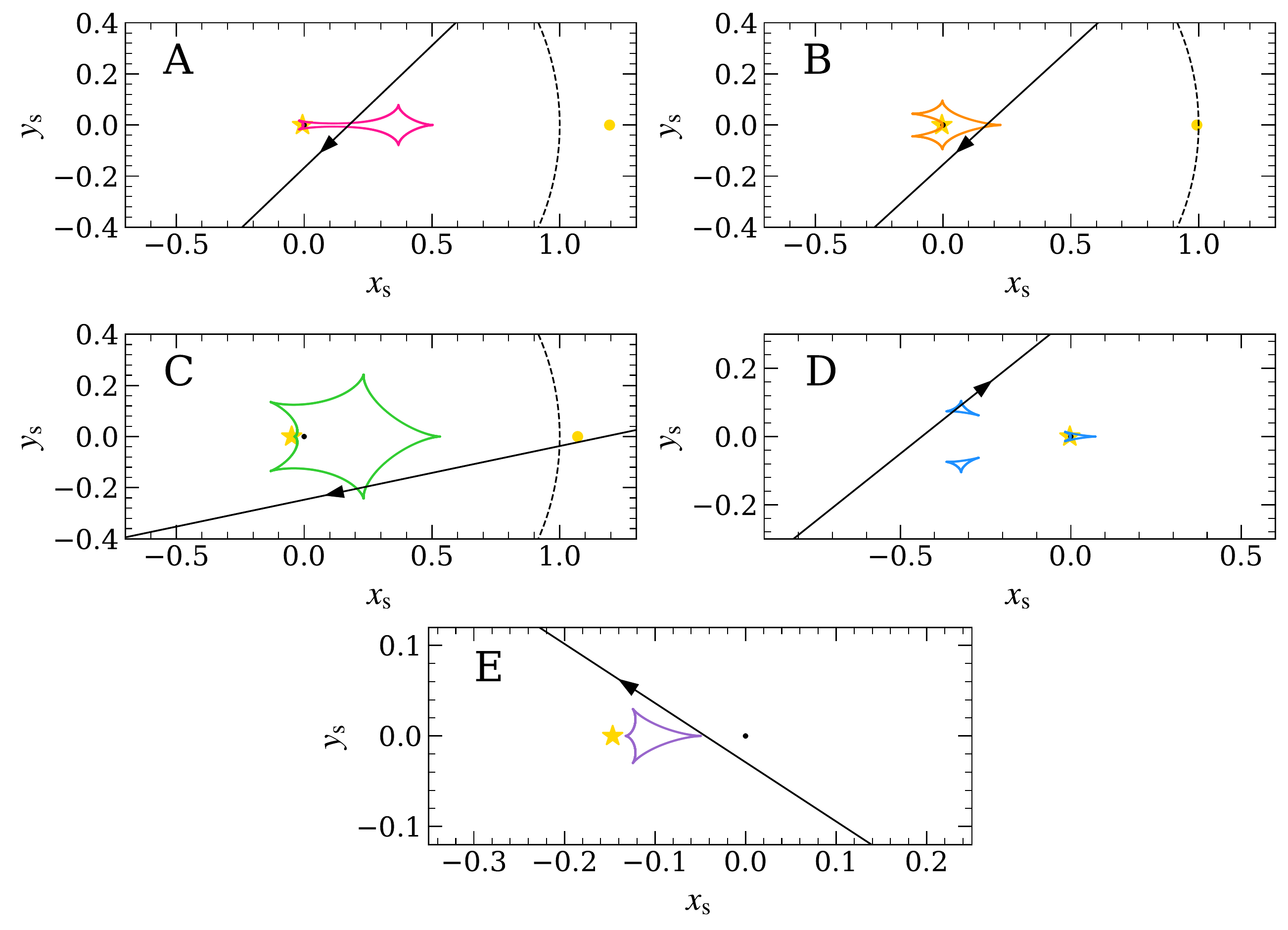}
    \caption{Geometries of the five different binary-lens models. In each panel, the caustic is color-coded to match the light curves in Figures \ref{lc1} and \ref{lc2}. The yellow dots represent the positions of the planet, and the yellow asterisks represent the positions of the host star. The black solid line is the trajectory of the source, and the arrow indicates the direction of the source motion. The axes are in units of the Einstein radius $\thetae$, and the black dashed line is the angular Einstein ring of the lens system.}
    \label{cau}
\end{figure}

\begin{figure}[htb] 
    \includegraphics[width=\columnwidth]{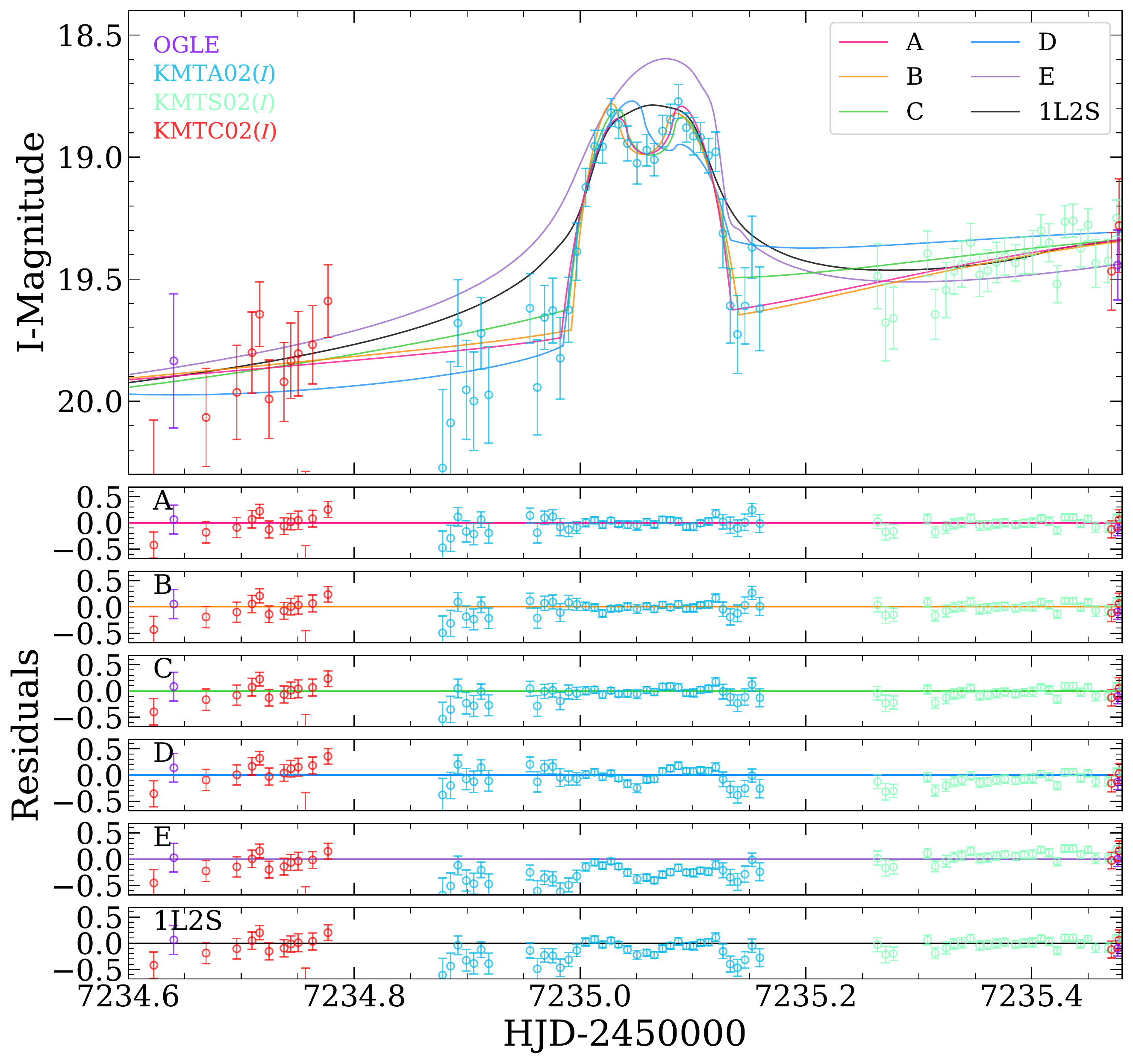}
    \caption{A zoom of the planetary anomaly region. The Symbols are the same as those in Figure \ref{lc1}.}
    \label{lc2}
\end{figure}

\begin{figure}[htbp]
    \centering
    \subfigure{
    \begin{minipage}{16.5cm}
    \centering
    \includegraphics[width=\columnwidth]{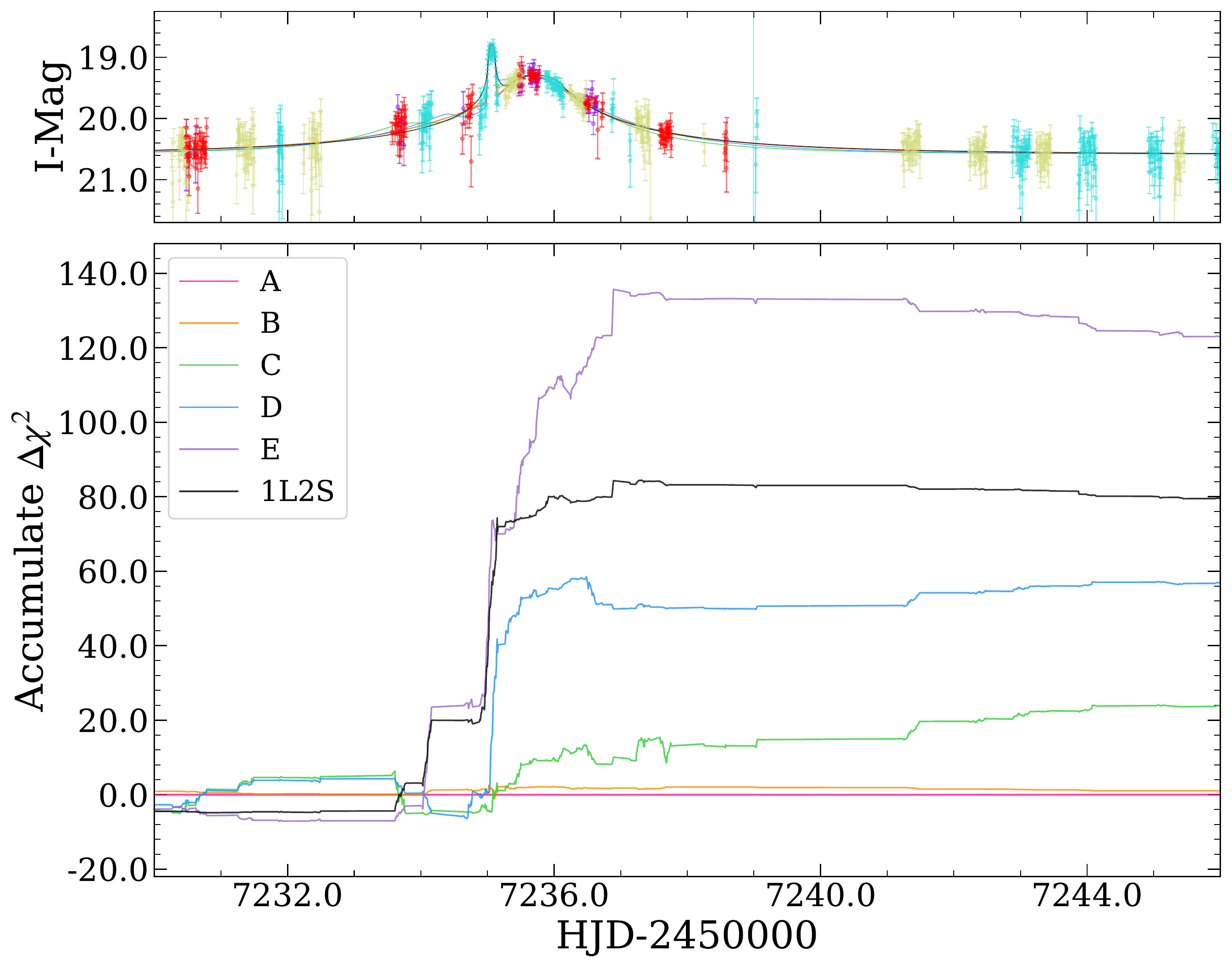}
    \end{minipage}
    }
    \caption{Cumulative distribution of $\chi^2$ differences for the 2L1S models and the 1L2S model compared to the 2L1S Model A ($\Delta\chi^2 = \chi^2_{\rm model} - \chi^2_{\rm A}$) as a function of time.}
    \label{chi2}
\end{figure}

\begin{figure}[htb] 
    \includegraphics[width=\columnwidth]{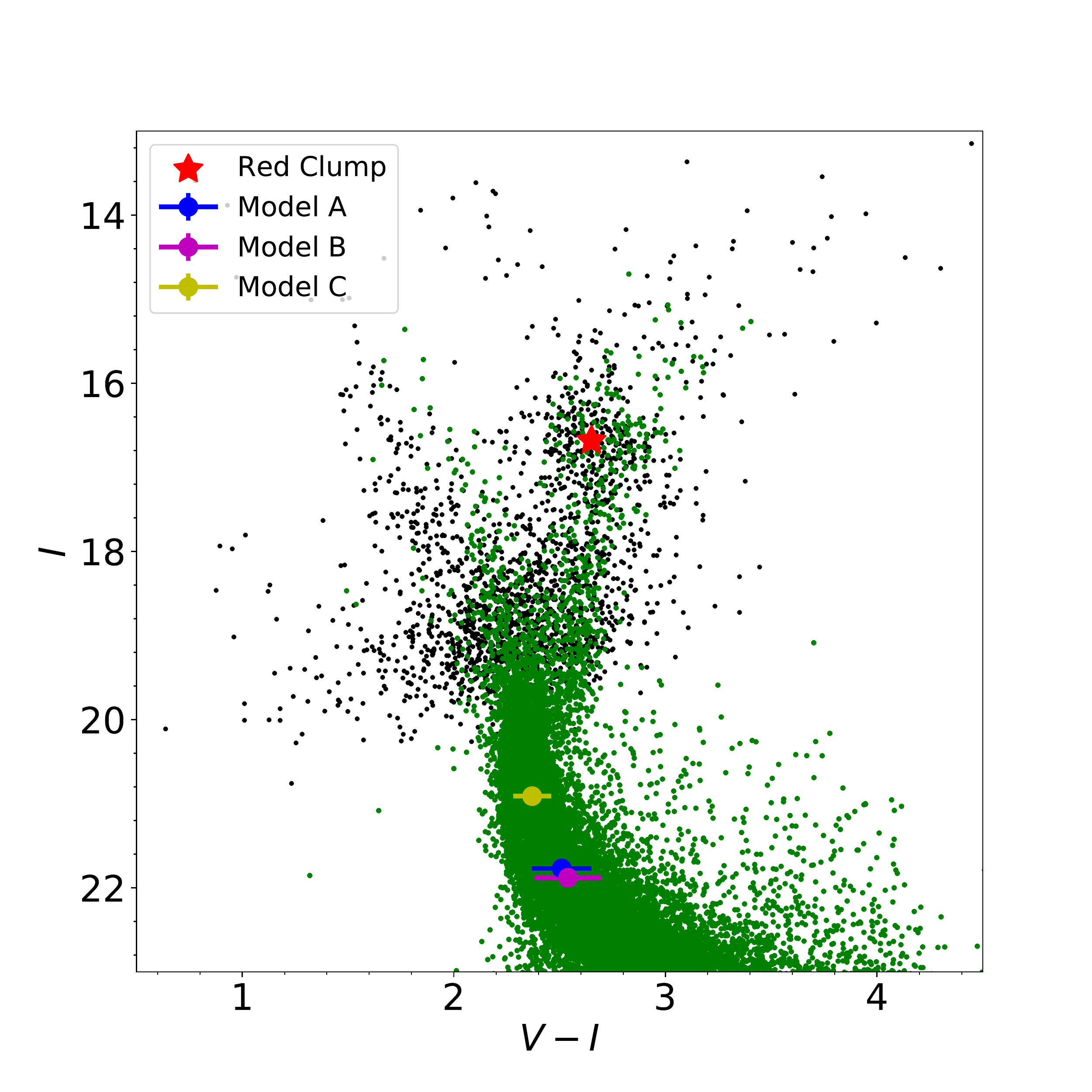}
    \caption{Color-magnitude diagram of a $2′ \times 2′$ square centered on  \event. The black dots show the stars from the OGLE catalog, which are roughly calibrated to the standard filter using the formula of \cite{OGLEIV}. The green dots show the HST CMD of \cite{HSTCMD} whose red-clump centroid is adjusted to OGLE's using the Holtzman field red-clump centroid of$(V - I, I)=(1.62, 15.15)$ \citep{MB07192}. The red asterisk shows the centroid of the red clump, and the blue, magenta and yellow dots represent the position of the source of different models.}
    \label{cmd}
\end{figure}

\begin{figure}[htb] 
    \centering
    \includegraphics[width=0.85\columnwidth]{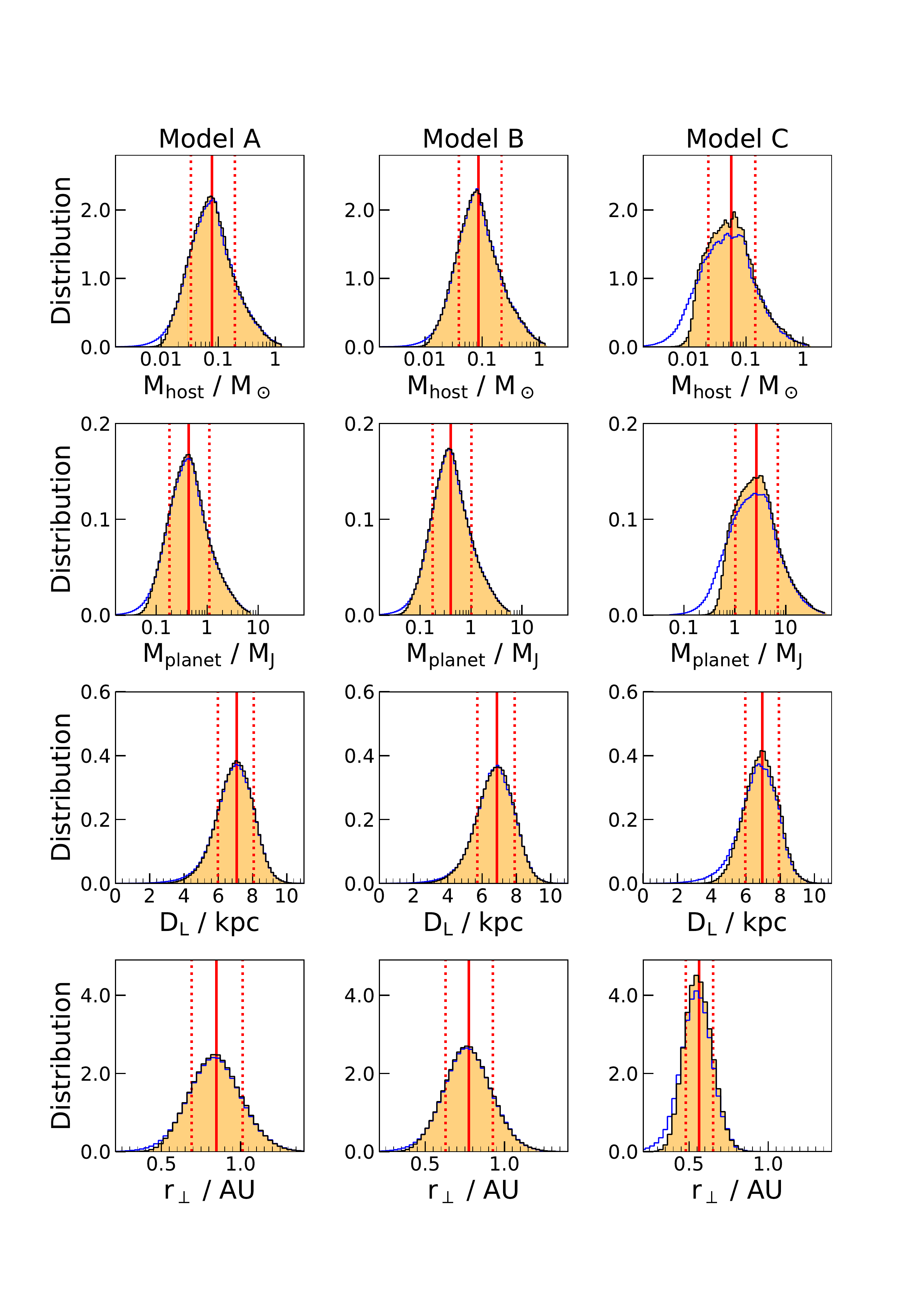}
    \caption{Bayesian posterior distributions of the lens host-mass $M_{\rm host}$, the lens distance $D_{\rm L}$, the planet mass $M_{\rm planet}$ and the projected planet-host separation $r_{\bot}$ for Models ``A'', ``B'' and ``C''. In each panel, the distribution marked in black color is obtained with $\alpha_{\rm pl} = -4.0$, while that marked in blue color are derived with $\alpha_{\rm pl} = 0.6$. The red solid vertical line and the two red dashed lines represent the median value and the 16th and 84th percentiles of the distribution obtained with $\alpha_{\rm pl} = -4.0$.}
    \label{fig:baye}
\end{figure}
